\documentclass{ws06-procs}

\begin{document}

\title{Star Formation Rate and Environment in the SDSS-DR4}

\author{G. Sorrentino}
\address{INAF-Osservatorio Astronomico di Capodimonte, 
         via Moiariello 16, 80131 - Napoli }  
\address{{\it VSTceN}, via Moiariello 16, 80131 - Napoli }

\author{A. Rifatto}
\address{INAF-Osservatorio Astronomico di Capodimonte, 
         via Moiariello 16, 80131 - Napoli } 
\maketitle
%
%
\abstracts{We  investigate the  environmental  dependence of  galaxies
with the star formation rate (SFR) in a complete volume limited sample
of 91566 galaxies,  in the redshift range $0.05 \le  z \le 0.095$, and
with  $M_r=-20.0$ (that  is M$^*$  +  1.45), selected  from the  Sloan
Digital  Sky Survey  Data  Release 4  (SDSS-DR4).  The environment  is
characterized by the local number  density of galaxies, defined by the
parameter $\Sigma_{N}(\frac{N}{\pi*r_{N}^{2}})$,  with N=5. We  find a
reletaion between the  distance of the nearest neighbour  and the SFR,
and  confirm  the general  trend  for  the  SFR of  decreasesing  with
increasing density.}
%
%
\section{Introduction}
There  are many  results  showing  that galaxy  properties  vary as  a function
of the environments. For example, one of the most fundamental correlation
between the properties  of galaxies and the environment in the  local universe 
is the  so-called morphology-density  relation in clusters  (e.g., Dressler 
1980; Treu  et al.  2003; Sorrentino  et al. 2006) and  its associated SFR -
density relation (e.g., Lewis  2002). SFR also depends  on the local  density
in groups and in the  field. For example, using the Early  Data Release (EDR)
of the Sloan  Digital sky  Survey (SDSS), Go\'{m}ez  et al. (2003) found that
the SFR - density relation also hold for field galaxies.  Moreover, Balogh et
al.   (2004) used both the Two-Degree  Field Galaxy Redshift Survey (2dFGRS;
Colless et al. 2001) and the SDSS to show that the SFR of  field  galaxies  is 
strongly  dependent on the local projected density. All these results  point 
to  the  existence  of  physical mechanisms that quench star formation as the
local density increases, from  the  field  to  groups  to  clusters.   This 
suggest  that  the morphology - density relation  is not  driven by processes 
that operate only in  extreme environments, such  as the core of  rich
clusters. In this  paper we attempt to gain  new insight into the  nature of
the previous relations, by studying how SFR depends on environment.
%
%
\section{The SDSS-DR4} 
The SDSS (York et al. 2000;  Abazajian et al. 2004) is a five-passband
({\itshape    $u'$,   $g'$,   $r'$,    $i'$,   $z'$})    imaging   and
medium-resolution  (R  $\simeq$  1800)  spectroscopic  survey  of  the
northern Galactic hemisphere. In  the DR4, the photometric area covers
6670 square  degrees, while the spectroscopic area  covers 4783 square
degrees,  providing  spectra  of  about  $10^{6}$  galaxies,  $10^{5}$
quasars, 30,000  stars and 30,000 serendipity targets  in the spectral
range $3800<\lambda<9200$  \AA, with a  rms redshift accuracy  of $30$
km~s$^{-1}$ to  an apparent  magnitude limit (Petrosian  magnitude) of
$r'=17.77$.\\  Spectroscopic   data  are  obtained  with   a  pair  of
multi-fiber spectrographs. Each  fiber has a diameter of  0.2 mm ($3"$
on the sky),  and adjacent fibers cannot be  located more closely than
55"  on the  sky ($\sim$  110 kpc  at $z$  = 0.1  with H$_0$  =  75 km
s$^{-1}$ Mpc$^{-1}$) during the same observation. In order to optimize
the placement of fibers on  individual plates, and as the placement of
plates  relative to  each other,  a tiling  method has  been developed
which allows  a sampling rate of  more than 92\% for  all targets. For
details         see          the         SDSS         web         site
(www.sdss.org/dr4/algorithms/tiling.html).\\  Data have  been obtained
from  the SDSS  database (http://www.sdss.org/DR4)  using  the CasJobs
facility (http://casjobs.sdss.org/casjobs/).
%
%
\section{Sample selection}
The SDSS-DR4  spectroscopic catalog is  magnitude-limited, complete to
the r-band  magnitude $m_{r}=17.77$.  In order to  avoid bias  we have
taken into account a complete volume-limited sample of galaxies in the
redshift range $0.05 \le z  \le 0.095$ and brighter than $M_r =-20.0$,
that is M$^*$ + 1.45. The  lower redshift limit is chosen with the aim
of minimizing the aperture bias  (Gom\'{e}z et al. 2003) caused by the
presence of large nearby galaxies, while the upper limit was estimated
through  Schmidt's  $V/V_{max}$ test.\\  
Our  initial sample  contains 91566  galaxies. For  each  galaxy, we  computed 
its r-band  absolute magnitude  corrected  for reddening  and  K  factor,  as
suggested  in Blanton et al. (2003).
\noindent
Galaxies with all  the seven emission lines nedeed  for the diagnostig
diagrams      (Kewley      et      al.      2001)      and      having
$I_{\lambda}/\sigma_{I_{\lambda}}  > 2$,  where  $I_{\lambda}$ is  the
emission  line flux  and $\sigma_{I_{\lambda}}$  its  uncertainty, are
classified as star forming  galaxies (SFGs), according to the criteria
adopted by Kewley  et al. (2001). In order to  avoid all the ambiguous
cases in  the AGN/SFG classification,  we removed those  sources whose
line  ratios  fall  close  to   the  border  line  of  the  diagnostic
diagrams. This was  done by keeping only those  sources for which part
of the error  bar associated to the logarithm  of the line-ratios lies
within  the theoretical uncertainty  of the  model ($\sigma_{mod}=0.1$
dex) in both $x$ and $y$ directions. 
The final sample consists of 11754 SFGs.
%
\begin{figure}[ht]
\centerline{\epsfxsize=3.9in\epsfbox{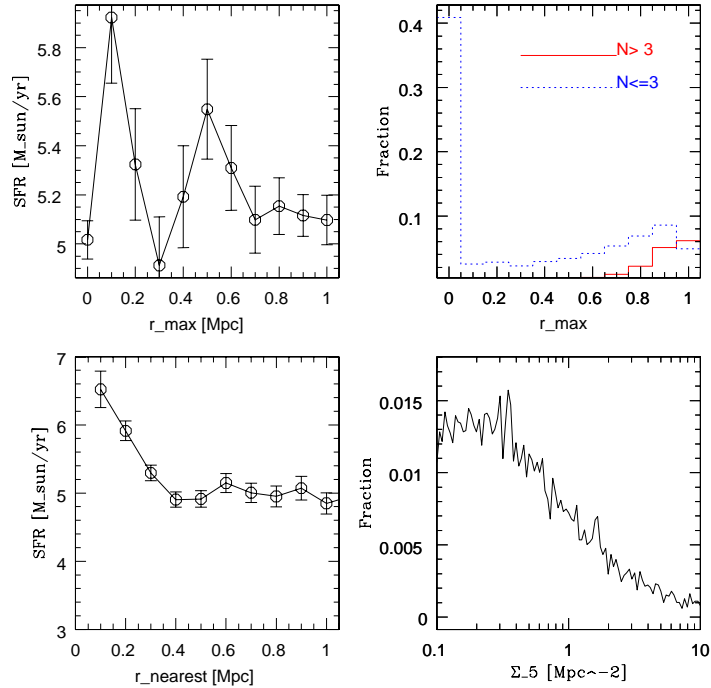}}   
\caption{ Top  left panel: distance  of the "most  distant neighbours"
(within  1 Mpc)  versus the  SFR. Top  right panel:  Neighbours number
distribution  with  distance.  Bottom  left  panel:  distance  of  the
"nearest  neighbours"  versus  the  SFR. Bottom  right  panel:  galaxy
density distribution.}
\end{figure}
%
%
\section{SFR and Environment}
For galaxies classified as star forming in the previous section, the SFR is
evalueted using the $H_{\alpha} line$, as in Hopkins et al. (2003):

\begin{eqnarray}
SFR(H_{\alpha}) & = & 7.9*10^{-42}L_{oc} \ \ M_{\odot}yr^{-1} \nonumber \\
                & = & (4\pi D_{L}^2 S_{H_{\alpha}})*
               \frac{10^{-0.4(r_{petro} - r_{fiber})}}{1.27*10^{34}} 
               \Bigg( \frac{S_{H_{\alpha}}/S_{H_{\beta}} }{2.86} \Bigg)^{2.114}
\end{eqnarray}
\noindent
The galaxy environment is defined by the $\Sigma_{5}$ density parameter,
corresponding to the density evalueted for the fifth nearest neighbours, within
1 Mpc and with $\Delta cz \le 1000 km/s$:
\begin{equation}
\Sigma_{5}=\frac{5}{\pi*r_{5}^{2}}
\end{equation}
The main  results are  summarized in Fig.~1.  In particular,  from the
bottom  left panel,  it is  evident that  the SFR  is enhanced  by the
presence  of a close  "nearest neighbour".  In fact,  the SFR  has its
maximum when  the distance of the  "nearest neighbour" is  0.1 Mpc. In
the range  from 0.1 to 0.4  Mpc it monotonically  decreases and after
0.4  Mpc the  SFR has  no variations  from the  mean value  of $\sim~5
M_{\odot}/year$.
\noindent
This result,  similar to that found  in previous works  using the data
from the 2dFGRS (Lambas et al. 2003, Sorrentino et al. 2003), confirms
that the presence of a  close companion can trigger the SFR. Moreover,
systems  having an  enanched  SFR are  formed  by a  little number  of
companions, as it is evident in Fig.1 - top right panel, where systems
with less than  three companions are separated from  systems with more
than three companions. For distances closer than 0.4 Mpc, we find only
poor systems.
\noindent
When we  look at  the large  scale environment, we  find a  result that
directly  remember  the  {\it morphology-density  relation}  (Dressler
1980). In fact,  from Fig.~1 - bottom right panel,  it is evident that
the fraction of SFGs decreases with the environment, following a trend
similar to  the "morphology-density relation"  for late-type galaxies.
Then,    this   result    suggests   the    esistence   of    a   {\it
morphology-activity-density  relation}, with three  parameters instead
of two (morphology-density).
%
%
\section{Conclusions}
In this  paper we  analized the environmental  dependence of SFR  in a
complete  volume-limited   sample  of   SFGs  in  the   SDSS-DR4.  The
environment is characterized by  the local number density of galaxies,
while the SFR is evalueted using the $H_{\alpha} line$.\\ 
Our findings can  be  summarized  in  the  following points:\\  
{\it  (i)}  SFR  is sensitive to  the local  galaxy density, in  such a way 
that galaxies show  higher levels  of star  formation  in low-density  than in 
high density environments.\\  
{\it (ii)} The  presence of a  close "nearest neighbour" enhances  the SFR,
which  reaches the maximum value  at the least distance  of 0.1 Mpc and  the
minimum, costant  value of $\sim~5 M_{\odot}/year$  at distances equal  or
greater  than 0.4  Mpc.\\ 
{\it (iii)} The environmental  properties of SFGs can be  related to a {\it
activity-density relation} in a  such way that  late-type galaxies are related
to the {\it morphology-density} relation.
%
%

\end{document}